\documentclass[12pt]{article}
\usepackage[dvips]{color}
\usepackage{graphicx}
\usepackage{amsthm,amssymb,amsmath,graphics,times}

\newcommand{\baseA}{\text{\sffamily A}}
\newcommand{\baseT}{\text{\sffamily T}}
\newcommand{\baseG}{\text{\sffamily G}}
\newcommand{\baseC}{\text{\sffamily C}}
\newcommand{\baseO}{\text{\sffamily O}}

\begin{document}

\centerline{\Large \sffamily Mismatch Repair Error Implies
Chargaff's Second Parity Rule}

\bigskip
\centerline{\sffamily Bo Deng\footnote{Department of Mathematics,
University of Nebraska-Lincoln, Lincoln, NE 68588. Email: {\tt
bdeng@math.unl.edu}}}

\bigskip
\noindent \textbf{Abstract: Chargaff's second parity rule (PR2)
holds empirically for most types of DNA that along single strands
of DNA the base contents are equal for complimentary bases,
$\baseA=\baseT, \baseG=\baseC$. A Markov chain model is
constructed to track the evolution of any single base position on
a given single strand of DNA whose organism is equipped with the
process of mismatch repair. Under the key assumptions that the
mismatch error rates primarily depend the complementarity and the
steric effect of the nucleotides and that the mismatch repairing
process itself makes strand recognition errors, the model shows
that the steady state probabilities for the base position to take
on one of the 4 nucleotide bases are equal for complimentary
bases. As a result, Chargaff's second parity rule is the
manifestation of the law of large numbers acting on the steady
state probabilities. More importantly, because the model pinpoints
mismatch repair as a basis of the rule, it is suitable for
experimental verification/falsification. At this point, existing
empirical survey supports the proposed mismatch-repair separation
between PR2 and non-PR2 genomes.}

\bigskip
\noindent{\bf Introduction.} Chargaff's parity rule with
$\baseA=\baseT, \baseG=\baseC$ for the nucleotide composition of DNA
(\cite{Char51}) is mechanistically explained by Watson-Crick's
double helix model of DNA (\cite{Wats53}) because base \baseA\ and
\baseT\ pair only with each other while base \baseG\ and \baseC\
pair only with each other along DNA's complementary double strands.
Chargaff and colleagues (\cite{Char68,Char69}) discovered a similar 
rule for single strands of DNA. For distinction, the former is
referred to as {\em Chargaff's Parity Rule} (PR) and the latter as
{\em Chargaff's Second Parity Rule} (PR2), respectively. Unlike PR,
PR2 is not an exact rule but a statistical one. More specifically,
according to a recent comprehensive test in \cite{Mitc06}, PR2 holds
for four of the five types of double stranded genomes: the archeal
chromosomes, the bacterial chromosomes, the eukaryotic chromosomes,
most double stranded viral genomes, but it does not apply to the
organellar (mitochondria and plastids) genomes nor to single
stranded viral genomes or any type of RNA genome. Another study
(\cite{Albr06}) also shows that the rule is length-dependent: for
those that PR2 does apply, the longer the strand is the smaller the
deviation from the rule becomes.


In contrast to PR, the {\em mechanistic} basis of PR2 is still an
open question (c.f. \cite{Lobr99b,Albr06}). Base inversion, base
substitution, base transposition are some of the suggested possible
causes (\cite{Albr06,Fick92,Sueo95,Lobr99a}). However, these
suggested mechanisms fail to make clear and definitive distinction
between those that obey the rule and those that do not. Various
mathematical models are also proposed to explain the rule. However,
they all led to a wrong prediction that single strands of a given
length would eventually evolve to exhibit the exact rule after long
enough time, which is inconsistent with the length-dependent
property of the rule. This defect was noted in (\cite{Lobr99b}).

Hence, any new model must make improvement in two fronts: its
applicability only to the PR2 genomes and its dependency on the
genomic size. The mathematical model we propose here seems to have
accomplished both by limiting its applicability (a) to those
organisms only with mismatch repair and (b) to the base evolution of
any arbitrary single position along any single strand of the
organism's DNA. The resulting model is a local Markov process for
single bases whose steady state probabilities together with their
manifestation via the law of large numbers give rise to the
length-dependent empirical PR2.


\bigskip
\noindent{\bf The Mathematical Model.} Let $m\{\text{\sffamily
B}_t\to\text{\sffamily B}_r\}$ denote the match or mismatch
probability of a replicative base $\text{\sffamily B}_r$ to a
template base $\text{\sffamily B}_t$. The mathematical model of DNA
replication with mismatch repair assumes the following.
\begin{quote}
{\bf Mismatch Repair Error (MRE) Model:}
\begin{enumerate}
\item[(a)] DNA replication makes nucleotide mismatch errors.
\item[(b)] At the moment of base replication,
the mismatch error occurs independent of the nucleotide base
position on the template single strand of DNA.
\item[(c)] The match probabilities satisfy the following
assumptions:
\begin{enumerate}
\item[i.] The match probabilities between complementary bases are equal with
respect to the complementary pairs:
\[
m\{\text{\baseA}\to\text{\baseT}\}=m\{\text{\baseT}\to\text{\baseA}\}\
\ \hbox{and}\ \
m\{\text{\baseG}\to\text{\baseC}\}=m\{\text{\baseC}\to\text{\baseG}\}
\]
\item[ii.] The match
probability of a base to its complementary or itself is the same as
that of the base's complementary to the complementary or itself:
\[
m\{\text{\baseA}\to\text{\baseA, or
 \baseT}\}=m\{\text{\baseT}\to\text{\baseA, or
 \baseT}\} \ \ \hbox{and}\ \
m\{\text{\baseG}\to\text{\baseG, or
\baseC}\}=m\{\text{\baseC}\to\text{\baseG, or \baseC}\}
\]
\item[iii.] The conditional mismatch probability of a pyrimidine to its
non-complementary purine equals that of its complementary purine to
the purine's non-complementary pyrimidine:
\[
m\{\text{\baseA}\to\text{\baseC}|\text{\baseA}\not\to\text{\baseT}\}=
m\{\text{\baseT}\to\text{\baseG}|\text{\baseT}\not\to\text{\baseA}\}
\]
\[
m\{\text{\baseG}\to\text{\baseT}|\text{\baseG}\not\to\text{\baseC}\}=
m\{\text{\baseC}\to\text{\baseA}|\text{\baseC}\not\to\text{\baseG}\}
\]
\end{enumerate}
\item[(d)] Mismatch repair
makes strand recognition errors independent of complementary bases,
and as a result it make base substitution errors.
\end{enumerate}
\end{quote}

Hypothesis (a) is self-evident, which applies to all DNA, with or
without mismatch repair. Hypothesis (b) is a localized and
symmetrical assumption that each base position is as critical or
ordinary as any other base position. It should be taken to be a {\em
primary} approximation of this aspect of the replication. This means
any assumption about global interactions along single strands of
DNA, such as the stem-loop hypothesis of \cite{Bell99}, or codon
position bias asymmetric assumption from \cite{Lobr99b}, can be
taken as secondary approximations or corrections for future
refinement of the model.

Hypotheses i,ii(c) implies that the mismatch probability of a base
to itself is the same as that of the base's complementary to the
complementary:
\[
m\{\text{\baseA}\to\text{\baseA}\}=m\{\text{\baseT}\to\text{\baseT}\}
\ \ \hbox{and}\ \
m\{\text{\baseG}\to\text{\baseG}\}=m\{\text{\baseC}\to\text{\baseC}\}
\]
Alternatively, Hypothesis ii(c) can also be stated in terms of the
mismatch conditional probabilities:
\[
m\{\text{\baseA}\to\text{\baseA}|\text{\baseA}\not\to\text{\baseT}\}=
m\{\text{\baseT}\to\text{\baseT}|\text{\baseT}\not\to\text{\baseA}\}
\]
\[
m\{\text{\baseG}\to\text{\baseG}|\text{\baseG}\not\to\text{\baseC}\}=
m\{\text{\baseC}\to\text{\baseC}|\text{\baseC}\not\to\text{\baseG}\}
\]
In other words, Hypotheses i,ii(c) are made mainly on the
complementarity of bases, which would be automatically true if DNA
were a binary code in either the \baseA\baseT-pair or the
\baseG\baseC-pair.

\begin{figure}
{\includegraphics[width=6.5in,height=1.75in]{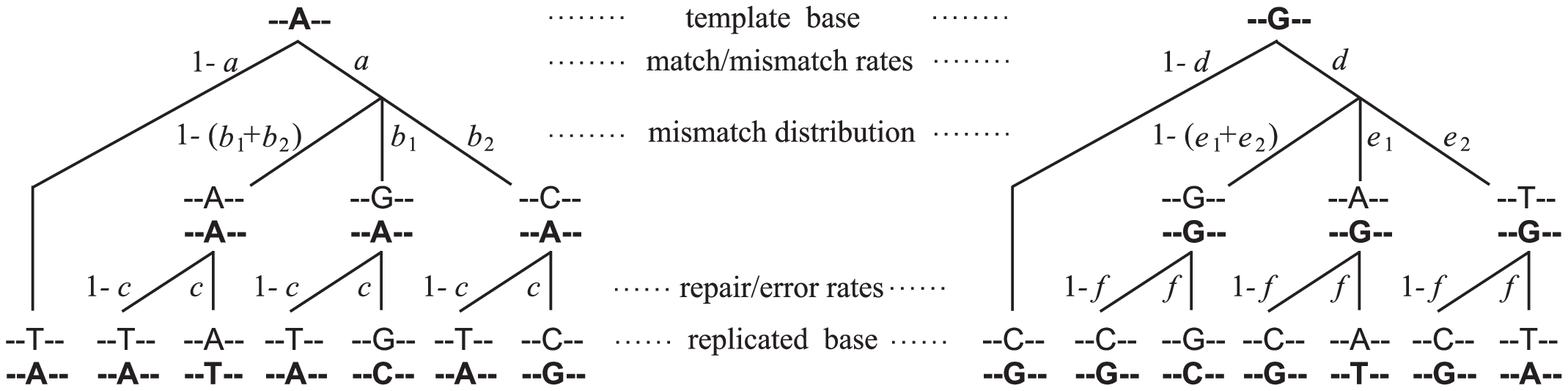}}

\caption{MRE charts for base \baseA\ and base \baseG. Parameters
$a,d$ denote the mismatch probabilities
$m\{\text{\baseA}\not\leftrightarrow\text{\baseT}\},
m\{\text{\baseG}\not\leftrightarrow\text{\baseC}\}$, respectively,
and $b_i,e_i$ denote the conditional mismatch probabilities. For
example,
$b_2=m\{\text{\baseA}\to\text{\baseC}|\text{\baseA}\not\to\text{\baseT}\}=
m\{\text{\baseT}\to\text{\baseG}|\text{\baseT}\not\to\text{\baseA}\}$
and similarly,
$e_2=m\{\text{\baseG}\to\text{\baseT}|\text{\baseG}\not\to\text{\baseC}\}=
m\{\text{\baseC}\to\text{\baseA}|\text{\baseC}\not\to\text{\baseG}\}$.}\label{figtransitiondiagram}
\end{figure}

On the other hand, however, Hypothesis iii(c) is made mainly on
the steric characteristics of the pyrimidines and purines, i.e.,
when an $\text{\baseA}\not\leftrightarrow\text{\baseT}$ (resp.
$\text{\baseG}\not\leftrightarrow\text{\baseC}$) mismatch occurs,
the conditional mismatch probability between \baseA$\to$\baseC\
(resp. \baseG$\to$\baseT) mismatch is the same as the
\baseT$\to$\baseG\ (resp. \baseC$\to$\baseA) mismatch. As a result
of Hypotheses i,ii,iii(c), the remaining conditional mismatch
probabilities are forced to satisfy:
\begin{equation}\label{eqConPro2}
\begin{array}{l}
m\{\text{\baseA}\to\text{\baseG}|\text{\baseA}\not\to\text{\baseT}\}=
m\{\text{\baseT}\to\text{\baseC}|\text{\baseT}\not\to\text{\baseA}\}
\\
m\{\text{\baseG}\to\text{\baseA}|\text{\baseG}\not\to\text{\baseC}\}=
m\{\text{\baseC}\to\text{\baseT}|\text{\baseC}\not\to\text{\baseG}\}
\end{array}
\end{equation}
More details can be found from the match/mismatch probability
diagrams of Fig.\ref{figtransitiondiagram}.

What separates our model from all others is Hypothesis (d). Although
similar assumptions such as base inversion, inverted transposition,
stem-loop substitution, codon position-bias were made for other
models, the precise and systematic mechanisms that made such
operations possible were either not known well enough or conjectured
too broadly for all types of DNA, to which we know by the result of
\cite{Mitc06} that PR2 does not always apply. As a result, none is
incorporated into the current model, nor is any repairing mechanism
outside the phase of replication such as excision repair to
spontaneous deamination of cytosine. In contrast, it appears that
most types of double-stranded chromosomes are equipped with mismatch
repair (\cite{Modr99}). More interestingly, mismatch repair in
mitochondria of any organism is not known to exist according to
\cite{Modr06}, even though other types of repair mechanisms may
exist (\cite{Bern06}) which do not negate the model under
consideration. This distinction between DNA having or not having
mismatch repair is consistent with the applicability of PR2
established in \cite{Mitc06}. As a result of this hypothesis, our
MRE model applies only to DNA types with mismatch repair satisfying
these four hypotheses.

\bigskip
\noindent{\bf The Result.} From a modeling perspective, Hypotheses
(c,d) imply that the MRE model is a Markov process model for any
arbitrary position of any single strand of DNA since the transition
probability at the position depends only on its current base. The
Markov model is illustrated in Fig.\ref{figtransitiondiagram},
referred to as the {\em mismatch repair error} chart, or MRE chart
for short. Take, for example, a nucleotide \baseA\ on a template
single strand of DNA as shown at the top of the left chart. With a
probability $0<1-a<1$, the process correctly replicates \baseA's
complementary base \baseT, showing down the left most branch, but
incorrectly with $0<a<1$ probability. Of that fraction of
mismatches, for a fraction of $0<b_1, b_2$ each the process
mismatches it with a \baseG\ or a \baseC, respectively, with
$b_1+b_2<1$. For the remaining $1-(b_1+b_2)$ fraction of mismatches,
an \baseA\ is mismatched to the original \baseA. These assumptions
are represented by the middle three branches.

For organisms which proof read and repair mismatches, the MRE chart
continues one level down to the bottom branches. In such cases,
Hypothesis (d) assumes that they do not always distinguish the
replicative strand from the template strand 100\% of the time, and
make strand recognition errors independent of the replicating bases
in question. Thus, for $0<c<1$ fraction of time, however
insignificant or small it may be at this point, the process mistakes
the replicative strand for the replicating template, and proceeds to
turn the original base \baseA\ into a base \baseT\ if the mismatched
pair is an \baseA-\baseA\ pair, with the hyphenated second base
position being the template base. Similarly, with a probability
$0<c<1$ each, a \baseG-\baseA\ pair is transformed to a
\baseG-\baseC\ pair and a \baseC-\baseA\ pair to a \baseC-\baseG\
pair, all changing the template base \baseA\ to a non-complementary
base. With $1-c$ probability, the process correctly identifies the
template strand, and proceeds to keep the original base \baseA\ in
the template strand when the replication is completed. This
particular illustration shows the case that the original mismatch
error is corrected by a base \baseT, which needs not to be the case.
In fact, what is only assumed and important is that with the $1-c$
probability the original base \baseA\ is {\em preserved} along the
template strand upon replication. The replicated new base along the
complementary strand may or may not be the complementary base of the
old base upon the completion of the mismatch repair. That is, the
most general depiction of the chart would replace the base \baseT\
by a place holder, \baseO, for any of the four bases.

Interchanging \baseA\ and \baseT\ and interchanging the parameters
$b_1,b_2$ in base \baseA's MRE chart gives rise to base \baseT's MRE
chart if the replicating base at the top starts with a \baseT.
Because of Hypotheses (c,d), all the probability distributions for
\baseT's chart are the same as base \baseA's chart except for the
conditional mismatch probabilities of Hypothesis iii(c) and the
consequence Eq.(\ref{eqConPro2}) of Hypothesis (c), which are
accounted for by interchanging $b_1$ and $b_2$. The transition
probabilities from base \baseT\ are listed on the second row of the
transition probability matrix $P$ below.

Similarly, the diagram on the right is base \baseG's MRE chart with
probability distributions in $d,e_1,e_2,f$, which may not be
respectively the same as $a,b_1,b_2,c$ for \baseA\ and \baseT. In
particular, $a$ and $d$ are very likely not to be the same, so are
for $b_i$ and $e_i$. According to Hypothesis (d), we have $c=f$. But
there will be no changes to the result if we keep $c,f$ distinct.
Also, interchanging \baseG\ and \baseC\ in \baseG's MRE chart gives
rise to \baseC's MRE chart.

The mismatch repair charts allow us to derive the transition
probabilities between bases one replicating generation a time. More
specifically, take the \baseA's MRE chart for an example. Let
$p\{\text{\baseA} \to \text{\baseA}\}$ be the probability that a
base \baseA\ from a template strand remains an \baseA\ upon
replication. Similarly, let $p\{\text{\baseA} \to \text{\baseT}\}$
be the the probability that a base \baseA\ from the template strand
is substituted by a \baseT\ upon replication, and similar notation
applies to $p\{\text{\baseA} \to \text{\baseG}\}, p\{\text{\baseA}
\to \text{\baseC}\}$. Then these probabilities can be tabulated from
probability distributions from the MRE chart as follows:
\begin{equation*}
\begin{split}
p\{\text{\baseA} \to \text{\baseT}\} &=a(1-(b_1+b_2))c\\
p\{\text{\baseA} \to \text{\baseG}\} &=ab_1c\\
p\{\text{\baseA} \to \text{\baseC}\} &=ab_2c\\
p\{\text{\baseA} \to \text{\baseA}\} &=1-p\{\text{\baseA} \to
\text{\baseT}\} -p\{\text{\baseA} \to \text{\baseG}\}-
p\{\text{\baseA} \to \text{\baseC}\} = 1-ac\\
\end{split}
\end{equation*}
The first expression, for example, is obtained by following from the
top of the chart down the direct branches leading to the
substituting base \baseT\ and multiplying all the probabilities
along the branches. The same for the transition probabilities from
\baseA\ to \baseG\ and \baseC\, respectively. The probability from
\baseA\ to \baseA\ can either be obtained by the formula above or by
summing all probabilities of the \baseA-to-\baseA\ branches (four
branches in all) of the chart. Exactly the same tabulation gives the
transition probabilities for all other bases.

Using matrix entry notation, we denote
\[
p_{11}=p\{\text{\baseA} \to \text{\baseA}\},\
p_{12}=p\{\text{\baseA} \to \text{\baseT}\},\
p_{13}=p\{\text{\baseA} \to \text{\baseG}\},\
p_{14}=p\{\text{\baseA} \to \text{\baseC}\}.
\]
That is, the row and column numbers, 1, 2, 3, 4, are in
correspondence with \baseA, \baseT, \baseG, \baseC, respectively.
Similar notation extends to other bases as well. As a result we
obtain the following transition probability matrix for our MRE
model:
\[
P=[p_{ij}]=
\left[\begin{array}{cccc} 1-ac & a(1-(b_1+b_2))c & ab_1c & ab_2c\\
a(1-(b_1+b_2))c & 1-ac & ab_2c & ab_1c\\
de_1f & de_2f & 1-df & d(1-(e_1+e_2))f\\
de_2f & de_1f & d(1-(e_1+e_2))f & 1-df
\end{array}\right]
\]
What follows are standard textbook properties of transition
probability matrixes of Markov chains (c.f.\cite{Meyn93}).

\begin{quote}
\begin{enumerate}
\leftskip -.5in
\item Denote the $n$th iterate of the transition
matrix by
\[
P^n=[p_{ij}^{(n)}].
\]
Then we know that it is again a transition probability matrix with
its entry $p_{ij}^{(n)}$ representing the probability of a base $i$
becoming a base $j$ after the $n$th generation of replication.
\item Because the transition matrix $P$ has all
positive entries (hence is irreducible and ergodic), by the
Perron-Frobenius Theorem, the limit $\lim_{n\to\infty}p_{ij}^{(n)}$
exists and the limit is independent of the initial base $i$:
$\lim_{n\to\infty}p_{ij}^{(n)}=\pi_j$, satisfying $0<\pi_j<1$ and
\[
\pi_1+\pi_2+\pi_3+\pi_4=1.
\]
In terms of DNA replication with mismatch repair, probability
$\pi_j$ is the {\em steady state probability} of finding base $j$ at
any base position of any single strand of DNA. This means,
regardless of the initial base $i$ at that position, after
sufficiently many generations of replication, the probability of
finding base $j$ at the position is $\pi_j$. We denote the four
steady state probabilities by $\pi_1=p\{\baseA\}, \pi_2=p\{\baseT\},
\pi_3=p\{\baseG\}, \pi_4=p\{\baseC\}$.
\item The transition matrix $P$ has $\lambda_1=1$ to be the largest
eigenvalue in magnitude and it is simple. Moreover, the steady state
probability vector $\pi=[\pi_1,\pi_2,\pi_3,\pi_4]$ is the only left
eigenvector not counting scalar multiple (or the transpose $\pi^{t}$
is the only right eigenvector of eigenvalue 1 for the transpose
matrix $P^{t}$). Because of this property, we can use the
eigenvector equation $\pi P=\pi$ to find $\pi$ explicitly as
\[
\pi=\frac{1}{2(a(b_1+b_2)c+d(e_1+e_2)f)}[d(e_1+e_2)f,\
d(e_1+e_2)f,\ a(b_1+b_2)c,\ a(b_1+b_2)c],
\]
with equal probabilities $\pi_1=\pi_2,\ \pi_3=\pi_4$ for
complementary bases. This, to be explained later, is the basis for
the empirical law of PR2.
\item By the Perron-Frobenius Theorem, the remaining eigenvalues of $P$
are less than 1 in magnitude. In this case, they can be explicitly
found as:
\begin{equation*}
\begin{array}{l}
\lambda_2 =1-a(b_1+b_2)c-d(e_1+e_2)f \\
\lambda_{3,4}=1-ac-df+\frac{1}{2}(a(b_1+b_2)c+d(e_1+e_2)f)
\pm\frac{1}{2}\sqrt{D},\ \hbox{ where }  \\
\qquad\quad
D=(-2ac+2df+a(b_1+b_2)c-d(e_1+e_2)f)^2\\
\qquad\qquad\qquad+4a(b_1-b_2)cd(e_1-e_2)f
\end{array}
\end{equation*}
It has the property that for any initial probability distribution
$q=[q_1,q_2,q_3,q_4]$ with $q_1+q_2+q_3+q_4=1$, $qP^n$ converges to
$\pi$ at a rate no greater than the order of $|\mu|^n$ with
$\mu=\max_{2\le i\le 4}\{|\lambda_i|\}<1$. This gives a temporal
estimate for the convergence rate to the steady state probability at
the $n$th generation of replication.
\end{enumerate}
\end{quote}

\leftskip 0in
\begin{figure}
\centerline{\includegraphics[width=3.5in,height=2.75in]{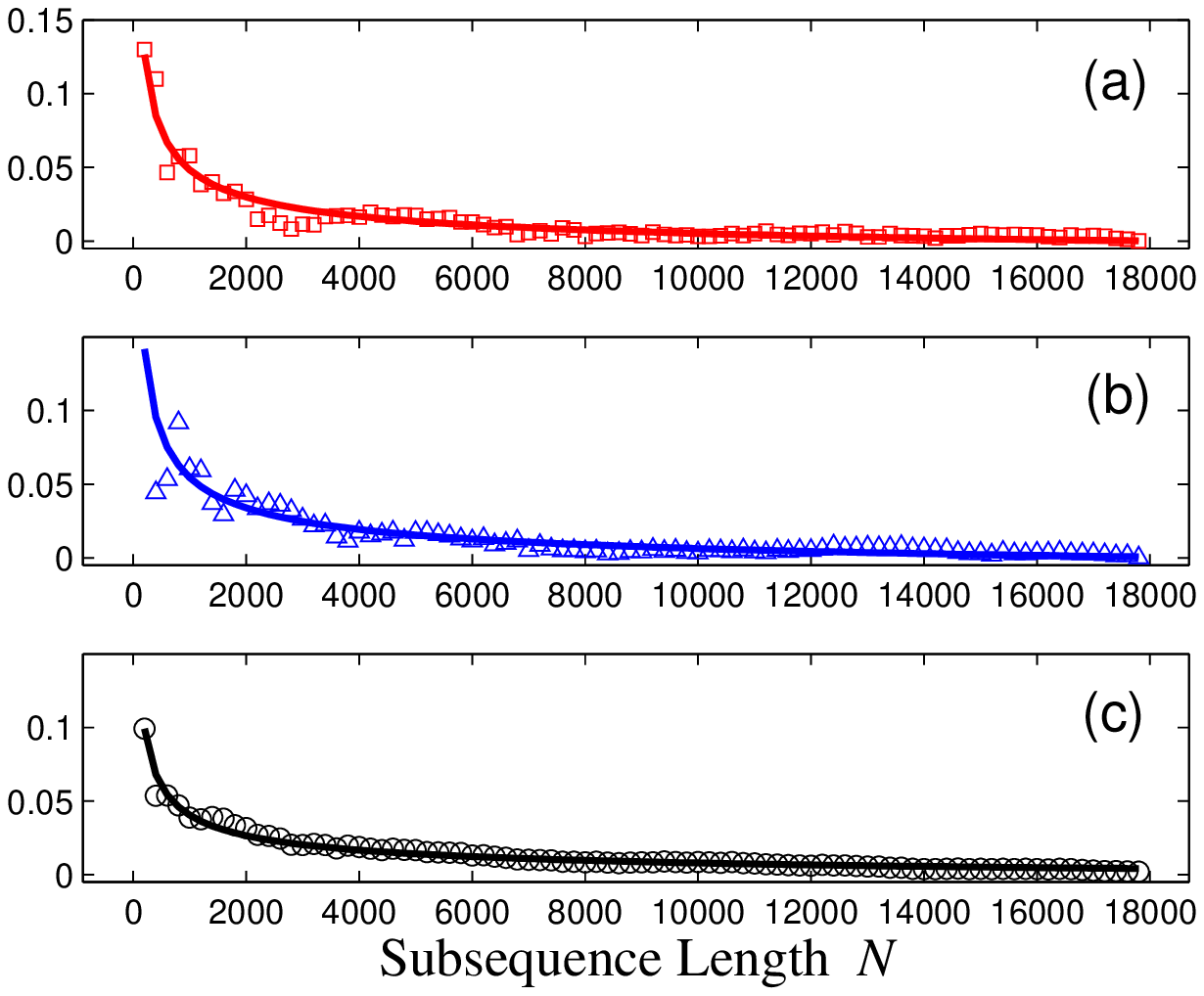}}
\caption{Homo sapiens chromosome 22 genomic contig
ref$|$NT$\underbar{\ \ }$113818.1$|$Hs22$\underbar{\ \
}$111537:1-17927 (from Human Genome Project) with length $L=17,927$
and ensemble frequency $F_L=(0.1993, 0.2007, 0.2917, 0.3083)$. (a)
In the $5'\to 3'$ direction, the sequence of ensemble frequencies
$F_N$ with length $N=1,2,\dots,L$ is generated. The convergence
error sequence,
$\sum_{i=\baseA,\baseT,\baseG,\baseC}|F_N(i)-F_L(i)|$, is plotted
against $N$ together with its best fit to the curve $a+b/\sqrt N$.
(b) The same plot except for the opposite $3'\to 5'$ direction. (c)
The same plot except (1) the data is averaged over 10 runs and (2)
each run is done for a random permutation of the original sequence.
}\label{figsimulation}
\end{figure}
%

\noindent We now conclude that Hypothesis (b), Property 2, and the
law of large numbers imply the empirical PR2. More specifically, let
$F_L\{i\}$ be the length-averaged frequency of base $i$, called the
{\em ensemble frequency}, then we must have
$\lim_{L\to\infty}F_L\{i\}=p\{i\}$ for $i=\baseA, \baseT, \baseG,
\baseC$. To see this, we borrow a prototypical explanation from coin
tossing: The steady state probabilities of the head and tail of a
coin can be approximated by tossing an ensemble of $L$ many
identical coins and counting the ensemble frequencies in which case
the greater the assemble size $L$ is the closer the ensemble
frequency to the steady state probability distribution becomes. In
fact, according to the law of large numbers (or the Central Limit
Theorem), the convergence rate has the order of $1/\sqrt L$.
Fig.\ref{figsimulation} gives a typical simulation of the
convergence for a short contig of the Human chromosome 22.
Simulations with longer sequences are not shown here because in
general the longer a sequence the better the convergence fit
becomes.

\medskip\noindent\textbf{Discussion.} It is interesting
to note a contrast between mismatch repair and other possible or
conjectured processes such as base inversion, transposition,
stem-loop substitution, codon position-bias asymmetry. For organisms
without mismatch repair, the MRE chart stops at the first or the
second branches with the template base well-entrenched in its
position. It is not clear what mechanistic operations these
processes adopt to {\em systematically} substitute the template
base. If such processes only subject the template base to a
sporadic, disorganized chance substitution as implied when such
systematic mechanisms are lacking, it is hard to see how a regular
and steady-state-like pattern such as PR2 can arise. It is also
important to note that the model starts with a single strand
separated from a double stranded mother DNA during replication and
ends with the completion of replication. This means that any damage
or repair that may take place before or after replication is not
relevant to the model at its current formulation. In other words, if
such types of damages and repairs are not the results of some
systematic, repetitive, and normal cellular functions, but rather
disorganized random events, then the lack of regular,
steady-state-like patterns in non-PR2 genome is not a contradiction
to the model.

On the surface of it, mismatch repair enhances replication fidelity
by enforcing PR. However, if the proposed model is correct, it will
imply that mismatch repair results in PR2. The question is then: is
PR2 also some sort of genomic fidelity that needs to be preserved?
If so, what is the evolutionary advantage of having it? A probable
answer can be found in the communication theory for DNA replication
proposed in \cite{Deng051} and it will be addressed elsewhere.

Past modeling attempts were not successful mathematically because
they did not make the critical distinction between the steady state
probability at individual base position and the ensemble frequency
of whole length along single strands of DNA. To a lesser extent,
they failed because they did not limit their scopes to different
types of DNA by pinpointing the mechanistic processes which
eventually but systematically led to the manifestation of PR2 we
observe today. These assessments motivated the model proposed here
which should be viewed as a first attempt to address both issues
simultaneously. Future refinements of the model should be done by
modifying and/or expanding the MRE hypotheses (a,b,c,d) in a manner
as concise as we did here.

Non-PR2 genomes can be used as an empirical control group to guard
against possible placebo effect of any proposed mechanism for PR2.
For example, the hypothesis of base inversion (\cite{Albr06}) can
be ruled out to be the cause of PR2 because it applies to both PR2
and non-PR2 genomes. Our proposed mismatch-repair mechanism, on
the other hand, has so far withstood this scrutiny by keeping the
empirical separation between PR2 and non-PR2 genomes.

Because of its conciseness, the model is suitable for experimental
test. More specifically, an experiment can start with many identical
double strands of DNA about 6,000 or more bp long (c.f.
Fig.\ref{figsimulation}) with an ensemble frequency disobeying PR2.
Divide them into two groups. Replicate one group by a replicator
with mismatch repair and replicate the other group by a replicator
without mismatch repair. After a {\em sufficiently} long period of
time, calculate the averaged ensemble frequency from each group and
compare. The experiment would be supportive of the model if the
ensemble frequency of the mismatch repairing replicator is closer to
PR2 than that of the null replicator. The model can also be rejected
if one can find a non-PR2 genome with mismatch repair.

\bigskip\bigskip\noindent
\textbf{Acknowledgement:} The author would like to thank Melissa
Wilson of Penn. State Univ., Dr. Modrich of Duke Univ., and Dr.
Lobry of Univ. Claude Bernard, France, for providing him with
various useful and important references. Special thanks to Dr.
Mitchell of Trinity College of Dublin for his invaluable comments on
the manuscript.

\end{document}